\documentclass{sf2a-conf}
\usepackage{graphicx}

\begin{document}
\TitreGlobal{SF2A 2007}
\title{Modeling of the multiwavelength emission of M\,87 with H.E.S.S. observations}


\author{J.-P. Lenain} \address{LUTH, Observatoire de Paris, CNRS, Universit{\'e} Paris Diderot; 5 Place Jules Janssen, 92190 Meudon, France\\ (jean-philippe.lenain@obspm.fr)}

\runningtitle{Modeling of the MWL emission of M\,87 with H.E.S.S.}

\setcounter{page}{1}

\index{Lenain J.-P.}

\maketitle

\begin{abstract}
  M\,87 is the first extragalactic source detected in the TeV range that is not a blazar. The large scale jet of M\,87 is not aligned with the line of sight. Modification of standard emission models of TeV blazars appears necessary to account for the $\gamma$-ray observations made by H.E.S.S. despite this misalignment.

  We present a new multi-blob synchrotron self-Compton model that deals explicitly with large viewing angles and moderate values of the Lorentz factor inferred from MHD simulations of jet formation.


\end{abstract}
%
\section{Introduction}

\begin{figure}[h]
  \centering
  \includegraphics[width=9cm]{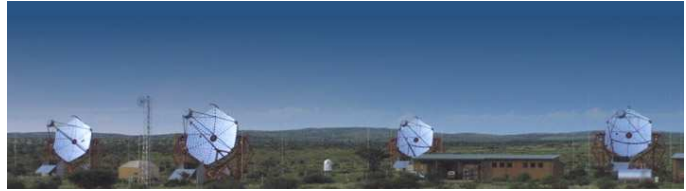}
  \caption{The H.E.S.S. array of 4 imaging atmospheric \v{C}erenkov telescopes in Namibia.}
  \label{fig:1}
\end{figure}

The H.E.S.S. collaboration operates an array of four 13\,m--imaging atmospheric \v{C}erenkov telescopes (Hinton 2004) located in Namibia at 1800\,m above sea level (see Fig.~\ref{fig:1}). The instrument measures $\gamma$-rays above a threshold of 100\,GeV up to several 10\,TeV by imaging the \v{C}erenkov light emitted by an air shower developing when a very high energy (VHE; $> 100$\,GeV) photon or particle enters the atmosphere.

The detection of M\,87 by HEGRA (Aharonian et al. 2003) and H.E.S.S. (Aharonian et al. 2006) has established M\,87 to be the first extragalactic object detected in the VHE range that does not belong to the blazar class, as its jet is not closely aligned with the line of sight. The multiwavelength emission of blazars is usually well described by leptonic radiative models, and requires high amplification due to a strong relativistic Doppler boosting.

We present here a multi-blob synchrotron self-Compton (SSC) emission model that explicitly deals with large angles to the line of sight to account for the recent observations of M\,87 in the TeV range made by H.E.S.S.

In Sect.~\ref{sect:classic} we present some results of a classical SSC model well suited to blazars and show the difficulty to directly apply it to M\,87. In Sect.~\ref{sect:multiblob}, we give some details on our multi-blob SSC model and its application to M\,87. Conclusions are given in Sect.~\ref{sect:concl}.

\section{Classical model}
\label{sect:classic}

We first present the results of a classical leptonic model (Katarzy{\'n}ski et al. 2001, 2003) applied to M\,87. This model describes the radiation transfer and emission by SSC processes in a single spherical blob of plasma moving at a relativistic speed along the jet axis. The blob is immersed in a uniform magnetic field and is assumed to be located inside the jet, close to the central supermassive black hole. The emission from radio to UV wavelengths is ascribed to an inhomogeneous conical extended jet. The absorption by the infrared extragalactic background light in the TeV range is also accounted for. However, since M\,87 is a nearby galaxy, its effect can be neglected. This model is well suited for modeling of blazars, namely when the velocity vector of the emitting zone is closely aligned with the line of sight.

\begin{figure}[h]
  \centering
  \includegraphics[width=9cm]{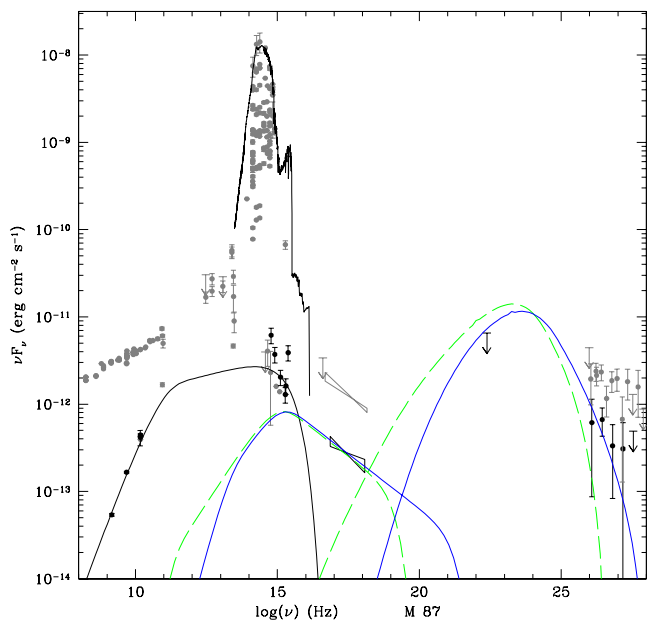}
  \caption{Spectral energy distribution of M\,87 with a classical SSC model.}
  \label{fig:2}
\end{figure}

To construct the spectral energy distribution (SED, see Fig.~\ref{fig:2}) of the core jet, simultaneous data are needed. Unfortunately, simultaneous data sets over a wide wavelength range are rare, so we carefully selected the data in the literature in order to have contemporaneous data or at least data representing the same state of activity. In particular, we assume here that the {\it Chandra} data taken in 2000 (Perlman et al. 2003) represent a low state of activity comparable to the one observed by H.E.S.S. in 2004 (Aharonian et al. 2006, black points in Fig.~\ref{fig:2}). The radio to optical/UV data are not simultaneous with the $\gamma$-ray data, but this is not problematic since these emissions mainly come from the extended jet, which has different properties than the zone emitting in VHE.

Figure~\ref{fig:2} represents the SED of M\,87 with two different classical models. The black curve in the radio band represents a solution for the extended jet. The black curve in the optical band is a modeling of the host galaxy. The green line represents the best solution of a single blob moving along the jet axis with a Doppler factor of 3 for a viewing angle of $19^\circ$, which is likely the greatest angle possible for the jet of M\,87 (Biretta et al. 1999). This solution does not describe correctly the emission at VHE. It would require much higher Doppler factor to describe them, which is not allowed because of the large value of the viewing angle.
The blue line represents the solution for a single blob moving along the line of sight in the jet formation zone with $\delta_b=8$. However this model is {\it ad hoc} since it is statistically unlikely that a single blob would move and emit exactly towards the line of sight.
As can be seen in Fig.~\ref{fig:2}, the classical SSC models hardly account for VHE data, thus it is necessary to modify these models to account for objects with misaligned jets.

\section{Multi-blob SSC model}
\label{sect:multiblob}

We present here a new jet emission model for AGNs based on the model of Katarzy{\'n}ski et al. (2001), for an emitting zone close to the central supermassive black hole. According to new results from general relativistic magnetohydrodynamic simulations of the jet formation (McKinney 2006), we can put constraints on macrophysics parameters in our model, such as the distance between the Alfv{\'e}n surface to the central supermassive black hole, the profile of the opening angle of the jet and the Lorentz factor of the material inside the jet.

\begin{figure}[h]
  \centering
  \includegraphics[width=9cm]{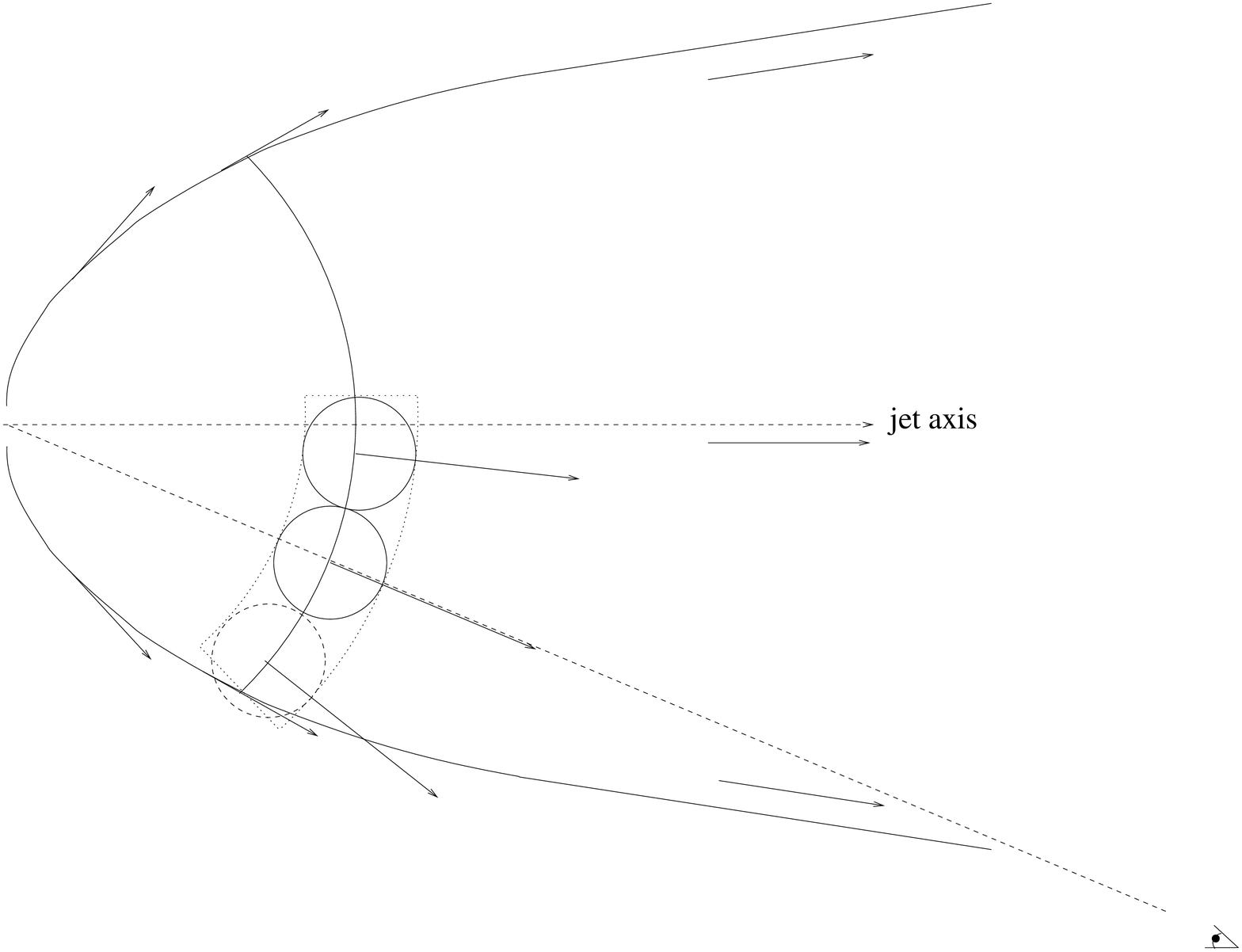}
  \caption{Geometric scheme of the multi-blob model.}
  \label{fig:3}
\end{figure}

The emitting zone is represented by a cap in the broadened zone at the base of the jet, located above the Alfv{\'e}n surface to let time to the acceleration process to take place (Lenain et al. 2007). This cap is filled with several blobs of plasma (see Fig.~\ref{fig:3}). In this way, differential Doppler boosting effect between the different blobs is accounted for.

The population of electrons thought to be responsible for the radio/optical emission through synchrotron radiation and X/$\gamma$-rays through inverse Compton process in leptonic models has a number density that is described by a broken power-law. In SSC models, these electrons radiate through synchrotron up to the X-rays in the case of M\,87, and then re-interact with the synchrotron photons by inverse Compton scattering, radiating up to VHE.

\begin{figure}[h]
  \centering
  \includegraphics[angle=-90,width=9cm]{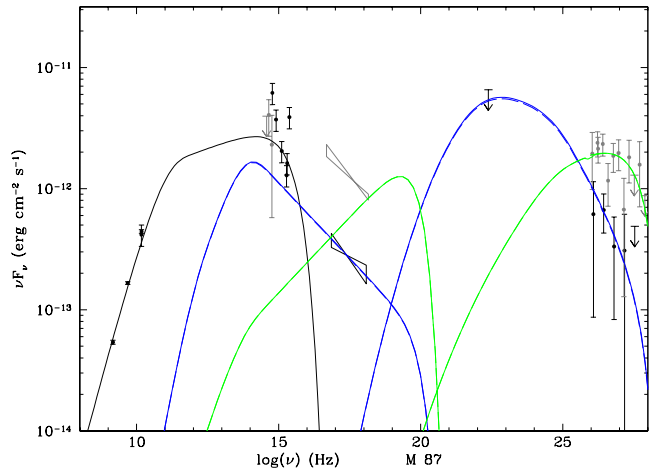}
  \caption{Spectral energy distribution of M\,87 with the multi-blob SSC model.}
  \label{fig:4}
\end{figure}

\begin{table}[h]
\caption{Parameters used in Fig.~\ref{fig:4}}
\label{table:1}
\centering
\begin{tabular}{c c c}
\hline\hline
& H.E.S.S. 2004 data (\it blue) & H.E.S.S. 2005 data (\it green)\\
\hline
$\Gamma_b$ & 10.0 & 10.0\\
$\theta$ & 15$^\circ$ & 15$^\circ$\\
$R_\mathrm{cap}$ [$r_g$] & 100.0 & 100.0\\
$B$ [G] & 0.01 & 0.01\\
$r_b$ [cm] & $2.8 \times 10^{14}$ & $8.0 \times 10^{13}$\\
$K_1$ [cm$^{-3}$] & $1.8 \times 10^4$ & $2.2 \times 10^4$\\
$n_1$ & 1.5 & 1.5\\
$n_2$ & 3.5 & 2.5\\
$\gamma_\mathrm{min}$ & $10^3$ & $10^3$\\
$\gamma_\mathrm{br}$ & $10^4$ & $10^4$\\
$\gamma_c$ & $10^7$ & $10^7$\\
\hline
\end{tabular}
\end{table}


We apply this multi-blob model to the multiwavelength emission of the jet of M\,87 to account for the recent observations at VHE by H.E.S.S. (see Fig.~\ref{fig:4}). The blue line shows the solution accounting for the low state of activity represented by the {\it Chandra} data of 2000 with the H.E.S.S. data of 2004, the corresponding parameters are described in column 2 of Table~\ref{table:1}, where $\Gamma_b$ is the Lorentz factor common to each blob, $\theta$ is the viewing angle with respect to the jet axis, $R_\mathrm{cap}$ is the distance between the central supermassive black hole and the cap of blobs, $r_g=G M_\mathrm{BH} / c^2$ is the scale length with $M_\mathrm{BH}$ the mass of the central black hole, $B$ is the magnetic field, $r_b$ stands for the individual radius of the blobs, the other parameters describe the population of electrons where $n_1$ and $n_2$ are the two indices of the broken power-law. The green line represents a solution accounting for the higher state of activity observed by H.E.S.S. in 2005 (gray points in Fig.~\ref{fig:4}). Unfortunately, no simultaneous X-ray data are available to constrain our model. However this illustrates the ability of the multi-blob model to generate spectra that are sufficiently hard in the VHE range to reproduce the most recent H.E.S.S. data. We obtain a characteristic length of the emitting zone of the order of $10^{14}$\,cm, comparable to the gravitational radius of this object, and interestingly with moderate values for the Lorentz factor of the blobs.

\section{Conclusions}
\label{sect:concl}

We propose an emission model for the internal jet of M\,87 giving solutions with small characteristic sizes of the emitting zone and moderate values for the Lorentz factor, agreeing with MHD simulations of jet formation. Slight modifications of standard SSC models for TeV blazars thus appears successful to account for the VHE data available up to now on M\,87.

We also applied the multi-blob model to other objects whose optical/X-ray jet is misaligned (PKS\,0521$-$36, 3C\,273 and Cen\,A, see Lenain 2007) to have a hint on their possible VHE flux.

\begin{acknowledgements}
  The author would like to thank Dr.~C.~Boisson and Dr.~H.~Sol for their invaluable help, and Dr.~A.~Zech for useful discussions.

  This research has made use of the NASA/IPAC Extragalactic Database (NED) which is operated by the Jet Propulsion Laboratory, California Institute of Technology, under contract with the National Aeronautics and Space Administration.
\end{acknowledgements} 



\end{document}